\documentclass[]{pasj01}
\begin{document} 
\Received{2015/09/08}%{yyyy/mm/dd}
\Accepted{2016/03/04}%{yyyy/mm/dd}
%\Published{yyyy/mm/dd}

\title{
An Effective Selection Method for Low-Mass Active Black Holes and First Spectroscopic Identification 
}

%%% begin:list of authors
% Do NOT capitalize all letters in "textsc".
\author{Tomoki \textsc{Morokuma}\altaffilmark{1,2}%
%\thanks{Example: Present Address is xxxxxxxxxx}
}
\altaffiltext{1}{Institute of Astronomy, Graduate School of Science, The University of Tokyo, 2-21-1 Osawa, Mitaka, Tokyo 181-0015, Japan}
\altaffiltext{2}{Kavli Institute for the Physics and Mathematics of the Universe (WPI), The University of Tokyo, 5-1-5 Kashiwanoha, Kashiwa, Chiba 277-8583, Japan}
\email{tmorokuma@ioa.s.u-tokyo.ac.jp}
\author{Nozomu \textsc{Tominaga}\altaffilmark{3,2}
\altaffiltext{3}{Department of Physics, Faculty of Science and Engineering, Konan University, 8-9-1 Okamoto, Kobe, Hyogo 658-8501, Japan}
}
\author{Masaomi \textsc{Tanaka}\altaffilmark{4,2}
\altaffiltext{4}{National Astronomical Observatory of Japan, 2-21-1 Osawa, Mitaka, Tokyo 181-8588, Japan}
}
\author{Naoki \textsc{Yasuda}\altaffilmark{2}}
\author{Hisanori \textsc{Furusawa}\altaffilmark{4}}
\author{Yuki \textsc{Taniguchi}\altaffilmark{1}}
\author{Takahiro \textsc{Kato}\altaffilmark{5}
\altaffiltext{5}{Department of Physics, Graduate School of Science, University of Tokyo, 7-3-1 Hongo, Bunkyo, Tokyo 113-0033, Japan
%Department of Physics, The University of Tokyo}
}
}
\author{Ji-an \textsc{Jiang}\altaffilmark{1}}
\author{Tohru \textsc{Nagao}\altaffilmark{6}
\altaffiltext{6}{Research Center for Space and Cosmic Evolution, Ehime University, Bunkyo-cho, Matsuyama 790-8577, Japan}
}
\author{Hanindyo \textsc{Kuncarayakti}\altaffilmark{7}
\altaffiltext{7}{Millennium Institute of Astrophysics, Casilla 36-D, Santiago, Chile}}
\author{Kana \textsc{Morokuma-Matsui}\altaffilmark{4}}
\author{Hiroyuki \textsc{Ikeda}\altaffilmark{4}}
\author{Sergei \textsc{Blinnikov}\altaffilmark{8,2,9}
\altaffiltext{8}{Institute for Theoretical and Experimental Physics (ITEP), Kurchatov Center, 117218 Moscow, Russia}
\altaffiltext{9}{Novosibirsk State University, 630090 Novosibirsk, Russia}
}
\author{Ken'ichi \textsc{Nomoto}\altaffilmark{2,10}
\altaffiltext{10}{Hamamatsu Professor}
\author{Mitsuru \textsc{Kokubo}\altaffilmark{1}}
\author{Mamoru \textsc{Doi}\altaffilmark{1}}
}
\author{Mitsuru \textsc{Kokubo}\altaffilmark{1,11}
\altaffiltext{11}{JSPS Fellow}
}
\author{Mamoru \textsc{Doi}\altaffilmark{1}}

%%% end:list of authors
\newcommand{\sersic}{S\'{e}rsic}

%% `\KeyWords{}' always has to be placed before `\maketitle'.
 %Do NOT move this preamble from here!

\maketitle

%%%%%%%%%%%%%%%%%%%%%%%%%%%%
\begin{abstract}
We present a new method to effectively select objects 
which may be low-mass active black holes (BHs) at galaxy centers 
using high-cadence optical imaging data, and 
our first spectroscopic identification of an active $2.7\times10^6$~M$_\odot$ BH at $z=0.164$. 
This active BH was originally selected due to its rapid optical variability, from a few hours to a day,
based on Subaru Hyper Suprime-Cam~(HSC) $g$-band imaging data taken with 1-hour cadence. 
Broad and narrow H$\alpha$ and many other emission lines are detected in our optical spectra taken with Subaru FOCAS,
and the BH mass is measured via the broad H$\alpha$ emission line width (1,880~km~s$^{-1}$) and 
luminosity ($4.2\times10^{40}$~erg~s$^{-1}$) after careful correction for the atmospheric absorption around 7,580-7,720\AA. 
We measure the Eddington ratio to be as low as 0.05, considerably smaller than those in a previous SDSS sample with similar BH mass and redshift, 
which indicates one of the strong potentials of our Subaru survey. 
The $g-r$ color and morphology of the extended component indicate that 
the host galaxy is a star-forming galaxy. 
We also show effectiveness of our variability selection for low-mass active BHs. 
\end{abstract}
%%%%%%%%%%%%%%%%%%%%%%%%%%%%

%%%%%%%%%%%%%%%%%%%%%%%%%%%%
%%%%%%%%%%%%%%%%%%%%%%%%%%%%
%%%%%%%%%%%%%%%%%%%%%%%%%%%%
\section{Introduction}
%%%%%%%%%%%%%%%%%%%%%%%%%%%%
%%%%%%%%%%%%%%%%%%%%%%%%%%%%
%%%%%%%%%%%%%%%%%%%%%%%%%%%%

Wide-field imaging and spectroscopic galaxy surveys 
such as Sloan Digital Sky Survey (SDSS; \cite{york2000}), 2dF \citep{croom2004}, and subsequent surveys 
have led to the discovery of a large number of quasars 
hosting active supermassive black holes (SMBHs), i.e., active galactic nuclei~(AGN), 
at their centers, even up to at a very high redshift ($z\sim7$, e.g., \cite{mortlock2011}). 
While bright AGN populations with large SMBHs have been discovered and studied 
in many ways over a wide range of wavelengths, 
observational studies of lower-mass BHs have been limited even in the local universe. 
Although every large SMBH should have experienced a phase of 
%growth from its seed stellar mass BHs,
%an observational link between stellar mass BHs and SMBHs has been elusive.
growth from its seed BH,
an observational link between seed BHs and SMBHs has been elusive.
A simple way to approach this problem is a survey to find such low-mass BHs. 
However, they are not easy to identify simply because of their faintness 
(i.e., Eddington luminosity $L_{\rm{Edd}}$ is proportional to a BH mass $M_{\rm{BH}}$, 
$L_{\rm{Edd}}\propto M_{\rm{BH}}$; see equation~\ref{eqn:variampl}). 

The huge number of the spectra taken in the SDSS project (\cite{greene2007b}; \cite{dong2012}) 
enable one to find 
%several tens 
%\textcolor{red}{a few hundreds} 
a few hundreds 
of low-mass ($<2\times10^6$~M$_\odot$ in their definitions) active BHs at $z<0.35$ 
by detailed fitting of the galaxy spectra (see also \cite{greene2004}; \cite{reines2013}). 
\citet{schramm2013} searched for X-ray sources in less massive galaxies 
with stellar masses of $<3\times10^{9}$~M$_\odot$ at $z<1$ 
utilizing optical-to-infrared spectral energy distributions (SEDs) and 
the deepest X-ray data in the Chandra Deep Field-South (CDF-S; \cite{xue2011})
and the Extended-CDF-S \citep{lehmer2005}.  
They found a $2\times10^5$~M$_\odot$ BH at $z=0.131$ 
based on the broad H$\alpha$ emission line %BH 
and two more plausible 
$\sim10^{5}$~M$_\odot$ BHs measured from the stellar bulge masses. 
In the local universe, in addition to 
the famous Seyfert 1 NGC~4395 with a $3.6\times10^{5}$~M$_\odot$ BH \citep{peterson2005} 
and other small BHs 
(\cite{barth2004}; \cite{reines2011}; \cite{reines2012}), 
\citet{baldassare2015} recently identified 
the smallest known active BH at a galaxy's center:
a $5\times10^4$~M$_\odot$ active BH 
with a low luminosity of $L_{\rm{bol}}=4\times10^{40}$~erg~s$^{-1}$ and 
a low Eddington ratio of $\sim0.01$ in a nearby dwarf galaxy 
at $z=0.0243$ \citep{reines2013}.

%In this \textcolor{red}{paper}, 
%{\it Letter}, 
In this paper, 
we introduce variability at optical wavelengths 
as an effective method %in \$\ref{sec:method}, 
which enables us to select AGN with lower BH masses, lower Eddington ratios, or at higher redshifts, 
compared to those found in the SDSS dataset. 
AGN in general show time variability all over the wavelength range. 
At UV-optical wavelengths, emission from an accretion disk is variable in time; 
this UV-optical variability has been (and will continue to be) used 
as one of the effective methods to select Type-1 AGN 
in many studies (\cite{devries2005}; 
\cite{sarajedini2006}; 
\cite{cohen2006}; 
\cite{morokuma2008a}; 
\cite{barth2014}; \cite{choi2014}). 
Considering 
the anti-correlation between luminosity and variability amplitude 
and 
the host galaxy light contamination, 
the variability information would be useful especially for lower-luminosity AGN 
if we could effectively extract variable components. 

First, we summarize our variability method in \$\ref{sec:method}. 
We describe our first 
high-cadence imaging survey (SHOOT; \cite{tominaga2016}) and 
optical spectroscopic identification 
of a low-mass active BH candidate in \$\ref{sec:obs}. 
%selected in our high-cadence imaging survey (SHOOT; \cite{tominaga2016}). 
Measurements of the black hole mass and the host galaxy properties are shown in \$\ref{sec:measurement}.
We discuss the effectiveness of our method and summarize the results in \$\ref{sec:summary}. 
We adopt the {\it WMAP5} cosmological model with $(H_{0}, \Omega_M, \Omega_\Lambda)=(70.5, 0.273, 0.726)$ \citep{komatsu2009}.
Galactic extinction is $A_g=0.075$ and $A_r=0.052$~mag \citep{schlafly2011} towards the AGN shown in this paper. 

%%%%%%%%%%%%%%%%%%%%%%%%%%%%%%%%%%%%%%%%%%%%%%%%%%%%%%%%%%%%%%%%%%%%%%%%%%%%%%%%
\section{Methodology for effective selection of low-mass active black holes}
\label{sec:method}
%%%%%%%%%%%%%%%%%%%%%%%%%%%%%%%%%%%%%%%%%%%%%%%%%%%%%%%%%%%%%%%%%%%%%%%%%%%%%%%%
UV-optical continuum emission of an accretion disk generally shows variability in time. 
Variability information has been used as an effective selection tool for quasars (e.g., \cite{hawkins1993}). 
It can be naively considered to be more effective for fainter populations because of 
an empirical anti-correlation between luminosity and variability amplitude \citep{vandenberk2004}. 
The dynamical time scale of an accretion disk emitting short-optical wavelength light, 
$t_{\rm{dyn}}=(R(\lambda)^3/GM)^{1/2}$, is as short as a few days for $10^6$~M$_\odot$ BHs. 
Thus, we suggest 
%\textcolor{blue}{
\begin{enumerate}
	\item[(i)]	rapid ($<1$~day) optical variability information 
	\item[(ii)]	at centers of galaxies 
	\item[(iii)]	obtained with deep and wide imaging data 
\end{enumerate}
%\vspace{-11mm}
as an effective method to select low-mass active BHs. 
NGC~4395, with a $3.6\times10^{5}$~M$_\odot$ BH, actually shows rapid ($<1$~day) variability in UV wavelengths 
even though the Eddington ratio is as low as $\sim10^{-3}$ \citep{peterson2005}. 
The timescale of variability is an important parameter
%The timescale of variations is an important parameter
in the selection process.
Luminous AGN with larger BH masses, quasars, vary most strongly over
long intervals, from months to years;
changes over periods of hours are generally small.
AGN with relativistic jets, known as blazars, do show 
large variations on short time scales and can be selected via variability in a short time scale 
(\cite{bauer2009}; \cite{ruan2012}; \cite{tanaka2014}) 
%(e.g., \cite{tanaka2014});
however, since they have a low number density \citep{ajello2012}, 
and can be recognized 
%easily 
by their apparent compactness and 
an SED characterized by synchrotron emission, 
they do not form a significant contaminant 
to samples of non-beamed AGN candidates 
showing rapid variability. 

This selection based on rapid variability is 
insensitive to redshift (since the time dilation effect only scales as $(1+z)$) 
except for inverse-square dimming with distance, which affects any selection method. 
In other words, variability on a short time scale,
unaccompanied by a relativistic jet, at any redshift, suggests 
that the origin is a small accretion disk around a small BH.
Host galaxy contamination does not have much influence on the variability detection 
because the sky is brighter than objects studied with 8-m class telescopes in optical imaging mode 
and dominates the observational errors. 
%while it does affect classical UV-optical color selection method very much 
%because the sky brightness dominates over the host galaxy light in the HSC data. 
On the other hand, 
other methods such as classical color selection, selection of low mass galaxies prior to X-ray selection, 
and spectral fitting for a large set of spectra, are affected by 
redshift and/or host galaxy contamination 
%\textcolor{blue}{
because a different redshift assumed gives a different rest-frame color and different stellar mass, 
and a blind search for active BHs can be done in a limited redshift range.
%}

%%%%%%%%%%%%%%%%%%%%%%%%%%%%
%%%%%%%%%%%%%%%%%%%%%%%%%%%%
%%%%%%%%%%%%%%%%%%%%%%%%%%%%
\section{Data \& Analysis}
%%%%%%%%%%%%%%%%%%%%%%%%%%%%
%%%%%%%%%%%%%%%%%%%%%%%%%%%%
%%%%%%%%%%%%%%%%%%%%%%%%%%%%
\label{sec:obs}

\subsection{Optical Transient Survey with Subaru Hyper Suprime-Cam}
We carried out HSC observations on July 2 (day 1) and 3 (day 2) in 2014 
for a field centered on 
(RA,Dec)=(16h32m12s.00, +35d02'57".1). 
%(RA,Dec)=(16h32m12s.00, +35d02'57".12). 
We adopted 1-hour cadence for finding rapid transients 
such as supernova shock breakouts 
(\cite{tominaga2011}; \cite{tominaga2016}; \cite{tanaka2016}). 
%(\cite{tominaga2011}; \cite{tominaga2014a}; 
%\cite{tominaga2014b}; \cite{tominaga2015a}; \cite{tominaga2015b}; \cite{tominaga2016}; \cite{tanaka2016}). 
In total, we obtained $g$-band images 
for 3 and 3 epochs on the days 1 and 2, respectively. 
Each epoch image consists of $5\times120$~sec exposures. 
The seeing was as good as 0.6~arcsec FWHM 
and the typical limiting magnitudes ($5\sigma$) are about 26.0~mag. 

The data was reduced with the standard HSC pipeline 
hscPipe version 3.6.1\footnote{A prototype is described in \cite{furusawa2010}.}, 
which is being developed based on the LSST pipeline (\cite{ivezic2008}; \cite{axelrod2010}). 
It provides packages for bias subtraction, flat fielding, astrometry, flux calibration, 
mosaicing, warping, coadding, source detection, and image subtraction. 
The astrometry and zeropoint magnitude determination  
are made relative to the Sloan Digital Sky Survey 
Data Release 8 (DR8; \cite{aihara2011}) 
with a 1.18~arcsec (7~pixel) aperture radius.  
%with a 2.36~arcsec (14~pixel) aperture diameter. 
We developed a quick image subtraction system with the pipeline and performed realtime transient 
finding in cooperation with an on-site data analysis 
system (\cite{furusawa2011}; \cite{furusawa2016}), which enables us to 
make catalogs of variable and transient sources 
right after observations (\cite{tominaga2014a}; \cite{tominaga2014b}; \cite{tominaga2015a}; \cite{tominaga2015b}). 
The realtime image subtraction method was applied to the stacked 600-sec exposure data 
and transient and variable sources were identified with the source detection code in hscPipe. 

The image subtraction procedure was developed based on the methodology shown in 
\citet{alard1998} and \citet{alard2000}. 
With the hscPipe, 
all the HSC images are consistently aligned with each other with common zeropoint magnitudes.
%All the HSC images are consistently aligned with each other in hscPipe. 
%{\bf MWR does not understand the following sentence:}
%Although the zeropoint magnitudes are common in the different HSC images, 
%the zeropoint magnitudes and the 
In the image subtraction procedure, point spread functions (PSFs) are 
matched from the reference image to the search images 
The image subtraction procedure is not perfect and sometimes provides 
artificial residuals in the subtracted images. 
One can judge
whether or not these residuals are real signals 
by measuring the shapes of the residuals (matching the PSF shape with the residuals). 
In the case for the object shown in this paper, 
we conclude that all the signals in the subtracted images are real. 

\begin{figure*}
 \begin{center}
  \includegraphics[angle=-90,width=12.5cm]{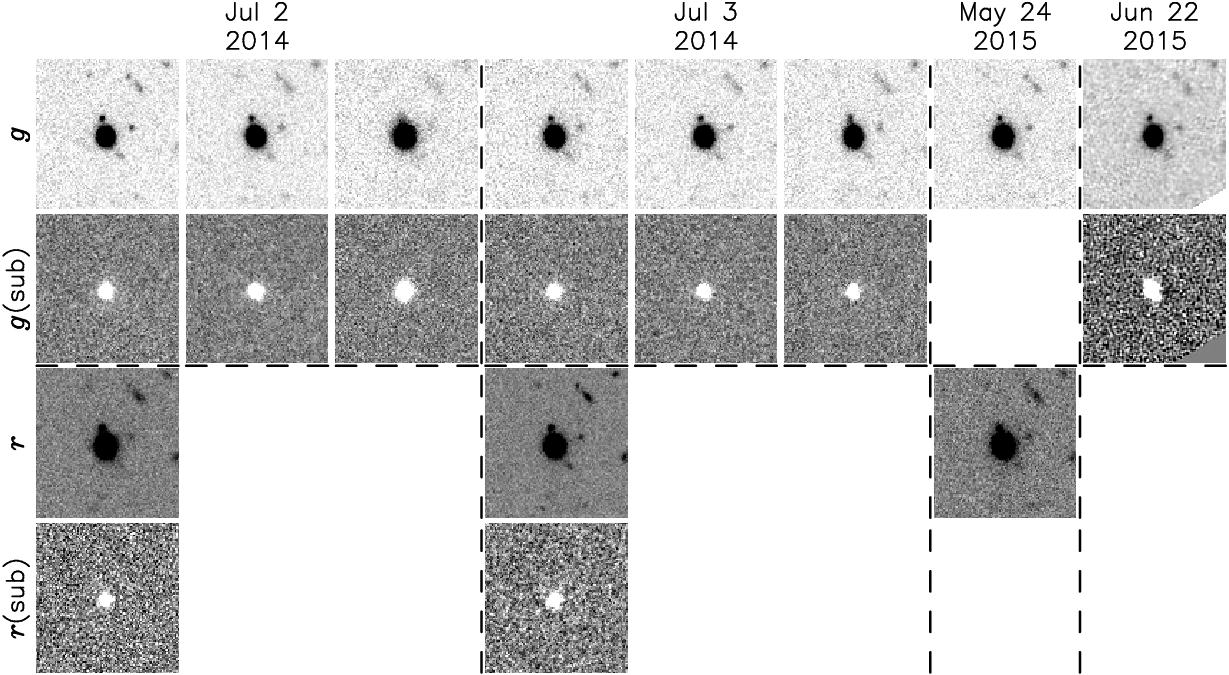}
  \includegraphics[angle=-90,width=12.5cm]{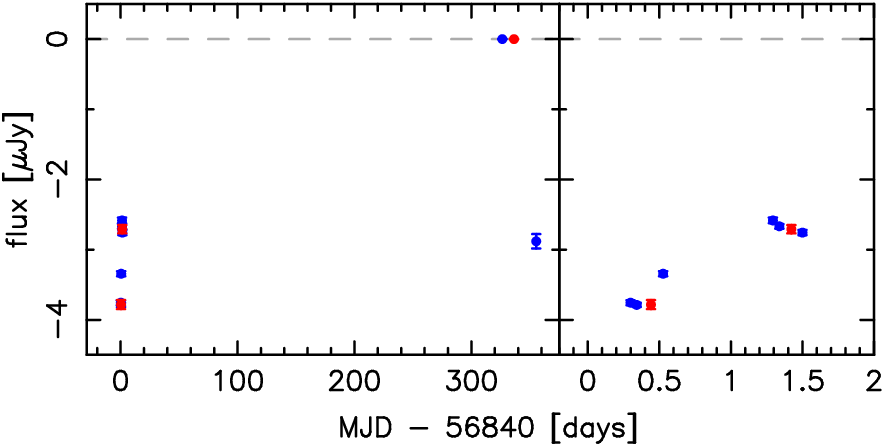}
 \end{center}
\caption{
(Top): Original and subtracted images in $g$- and $r$-bands taken with HSC and FOCAS. 
The field size is $100\times100$~HSC pixel corresponding to $16.9\times16.9$ arcsec$^2$. North is up and east is left. 
The image subtraction is done for each image with the HSC image taken on May 24, 2015 being the reference image in the both bands.
The original pixel scale of FOCAS is different from that of HSC. 
We matched the FOCAS data to the HSC data on the WCS basis only for this figure 
and the pixel scales of the images here are the same.
(Bottom): The entire light curves in $g$- and $r$-bands are shown in the left panel in blue and red circles, respectively. 
The $r$-band photometric point in the reference image (zero flux) in the left panel is shifted by 10 days for readers to see easily. 
These photometries are done 
in the subtracted images with references images taken on May 24, 2015. 
Light curves on Days 1 and 2 are magnified in the right panel. 
}
\label{fig:lcimage}
\end{figure*}

With this 2-night imaging data, we found a rapidly variable source at the center of an apparently small galaxy,
at (RA, Dec)=(16h33m56s.20, $+$35:13:39.3),
which is a good candidate of a low-mass BH. 
%at (RA, Dec)=(16h33m56s.204, $+$35:13:39.28). 
In addition to the discovery data, we took 1-epoch ($5\times120$~sec) $r$-band data on each night 
and $g$- and $r$-band data on May 24, 2015. 
The HSC data properties and the light curve of this object 
are summarized in Table~\ref{tab:lc} and Figure~\ref{fig:lcimage}. 
We note that no radio or X-ray counterparts are detected 
in the Faint Images of the Radio Sky at Twenty-Centimeters (FIRST; \cite{white1997}) 
or any of X-ray databases around this object, respectively. 

%\subsection{Follow-Up \textcolor{red}{Observations} with Subaru FOCAS}
\subsection{Follow-Up Observations with Subaru FOCAS}
\begin{figure*}
 \begin{center}
	\includegraphics[angle=-90,width=11cm]{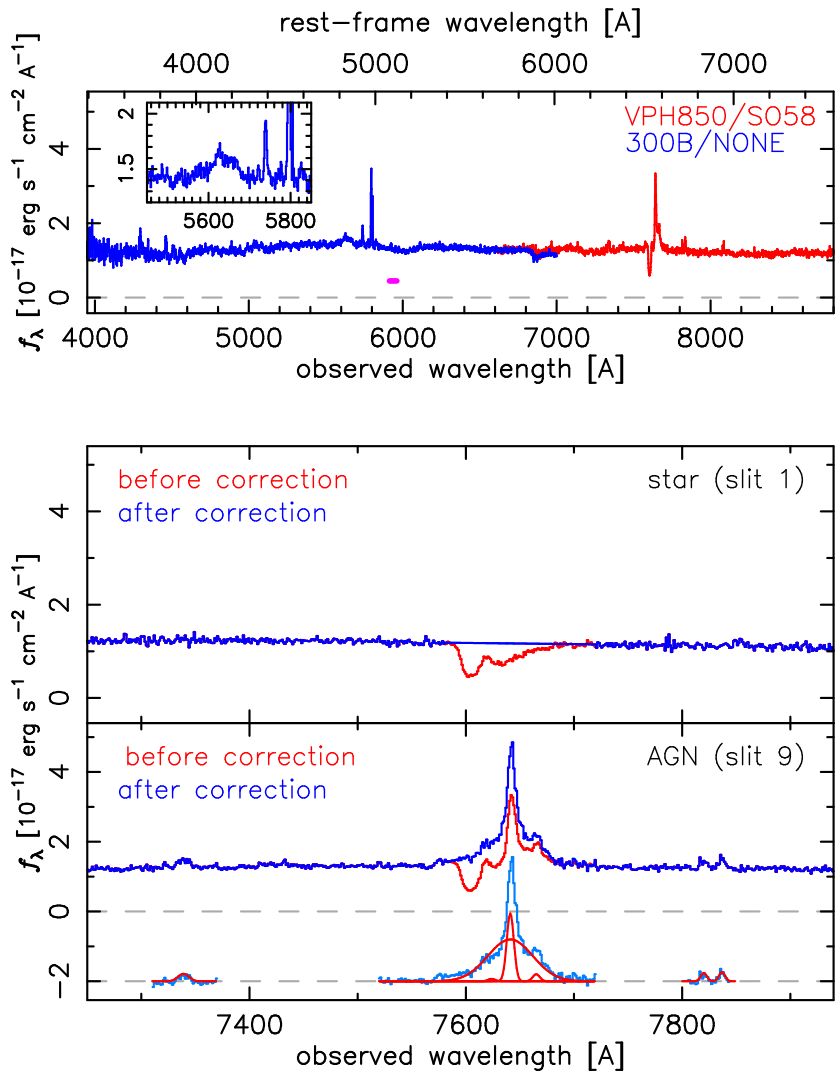} 
\end{center}
\caption{
Reduced FOCAS spectra. 
(top) Entire spectra shown in 
blue for 300B grism and non-filter and 
red for VPH850 grism and SO58 filter. 
AGN Flux density at rest-frame 5100\AA\ estimated from the broad H$\alpha$ luminosity is indicated in magenta. 
The broad H$\beta$ and [OIII] wavelength region is magnified in the inset. 
(middle) View around the [OI], H$\alpha$, [NII], and [SII] wavelength region. 
Spectra without and with atmospheric absorption 
(red and blue, respectively) for the star used for the correction and the target AGN. 
Continuum-subtracted spectra and fitted gaussians are shown in light-blue and red, respectively.
}\label{fig:spectrum}
\end{figure*}

We took $1200~\rm{sec}\times6$ optical spectra of this object 
with Faint Object Camera And Spectrograph~(FOCAS; \cite{kashikawa2002}) 
on the 8.2-m Subaru telescope on June 22, 2015. 
The slitmask was made for this region and each slit width was 1.0~arcsec. 
The position angle of the slits was 30~deg and 
the atmospheric dispersion corrector was used. 
Seeing was stable and as good as 0.5~arcsec FWHM. 
We adopted the $2\times1$ (spatial and dispersion; 0.208~arcsec~pixel$^{-1}$ and 1.19~\AA\ pixel$^{-1}$) binning mode. 
The first three of the 6 exposures were taken with the VPH850 grism and SO58 order-sort filter 
(giving $\lambda>5800$~\AA\ spectrum) 
while the latter three were taken with the 300B grism without any order-sort filters 
(giving $3500<\lambda<7000$~\AA\ spectrum without 2nd-order contamination). 
The nominal spectral resolutions were $R\sim600$ and $\sim400$ for these two observing modes, respectively. 
We note that the resulting spectral resolutions ($R\sim1,200$ and $\sim800$)
were better than these specifications because the image sizes 
were smaller than the slit widths. 
Flux calibration was done using the standard star Feige~110 
by utilizing the Hubble Space Telescope data (CALSPEC\footnote{http://www.stsci.edu/hst/observatory/crds/calspec.html}). 
The data reduction was carrired out first in a standard manner without giving any special attention  
to the atmospheric absorption feature around 7,580-7,720\AA. 
This is partly because very high-order polynomial fitting is required to 
correctly evaluate the transmission efficiency in this wavelength range;
and, more important, the standard star observation and object observation 
were done separately in time 
at a different airmass.
It would be more accurate to use simultaneous observational data for the correction,
and so 
we subsequently corrected this absorption feature as described in \S\ref{sec:aac}. 
The reduced spectra before the correction are shown in the top panel of Figure~\ref{fig:spectrum}. 

We also took $300$-sec $g$-band imaging data in $2\times2$~binning mode (0.208~arcsec~pixel$^{-1}$), 
which enables us to calibrate the flux more accurately. 
Image subtraction is applied to these FOCAS image after matching the HSC reference image to the FOCAS image 
because the pixel sampling of FOCAS is larger than that of HSC. 
Photometry for the data is also shown in Table~\ref{tab:lc} and Figure~\ref{fig:lcimage}. 
%Photometry for these data are also shown in Table~\ref{tab:lc} and Figure~\ref{fig:lcimage}. 

%%%%%%%%%%%%%%%%%%%%%%%%%%%%
%%%%%%%%%%%%%%%%%%%%%%%%%%%%
%%%%%%%%%%%%%%%%%%%%%%%%%%%%
%\section{Measurements of Emission Lines and Black Hole Mass}
\section{Measurements of Emission Lines, Black Hole Mass, and Host Galaxy Properties}
%\section{\color{red}{Measurements of Emission Lines, Black Hole Mass, and Host Galaxy Properties}}
\label{sec:measurement}
%%%%%%%%%%%%%%%%%%%%%%%%%%%%
%%%%%%%%%%%%%%%%%%%%%%%%%%%%
%%%%%%%%%%%%%%%%%%%%%%%%%%%%

%%%%%%%%%%%%%%%%%%%%%%
\subsection{Measurements of Emission Line Fluxes and Widths}
%\label{sec:measurement}
%%%%%%%%%%%%%%%%%%%%%%

In the FOCAS spectra, several emission lines are clearly detected and 
the line identification is secure.
However, the H$\alpha$ and [NII] emission line region, which is the most important 
wavelength region for our purpose, is unfortunately much affected by the strong 
atmospheric absorption around $7,580$-$7,720$~\AA. 
Therefore, we make a correction for this absorption and 
evaluate the error 
it might cause in 
our measurement of the BH mass. 

%%%%%%
\subsubsection{Atmospheric Absorption Correction}
\label{sec:aac}
%%%%%%

The sky was very clear and stable during the FOCAS observations, which was verified by the CFHT 
SkyProbe\footnote{http://www.cfht.hawaii.edu/cgi-bin/elixir/skyprobe.pl?plot\&mcal\_20150622.png}.
We consider the sky transmission (absorption) uniform within the 6~arcmin FOCAS field-of-view. 

In the slitmask, 1.0~arcsec slits were cut for a few stars as 
well as for the target AGN. 
One of the stars in the same slitmask is a G-type star,
based on the SDSS color and the FOCAS spectrum,
without any strong absorption or emission features
around the atmospheric absorption feature.
We estimated the absorption fraction by linearly (in the $\lambda$-$f_{\lambda}$ plane) 
fitting the adjacent continuum of the star spectrum (7550-7580\AA\ and 7750-7800\AA) 
and divided the AGN spectrum by this fraction. 

This procedure requires accurate and consistent wavelength calibration between the star and AGN spectra. 
We examined the positions of the star and AGN in the slits during the exposures. 
We note that the data was taken at a relatively high elevation 
and the change of the airmasses during the exposures was not so large, from 1.05 to 1.24. 
We found the slitmask was well aligned to every object 
and the difference of the positions to the center of the slits is conservatively evaluated to be 
smaller than 0.5~pixel, corresponding to 0.6~\AA. 
Then, we iterated the emission line fitting (described in \$\ref{sec:emlinemeasure}) 
by shifting the wavelengths of the reference star 
($-1.0$ to $+1.0$ pixel; i.e., $-1.2$~\AA\ to $+1.2$~\AA) 
and estimated a systematic error in the measurements of emission line widths and fluxes. 
We found that the effect of the wavelength difference is smaller 
than the scatter of the BH mass-to-H$\alpha$ properties
and does not affect our result very much.
The obtained systematic errors on the broad H$\alpha$ width, luminosity, and BH mass 
are about 10\%, 10\%, and 20\%, respectively. 
We note that the systematic error on the final BH mass is smaller than 
the equation used for calculating the BH mass from the H$\alpha$ properties. 
We below use the original AGN spectrum without this artificial offset and show only the statistical errors. 

%%%%%%%%%%%%%%%%%%%%%%%%%%%%%%%%%%%%%%%%%%%%%%%
\subsubsection{Emission Line Measurements}
\label{sec:emlinemeasure}
%%%%%%%%%%%%%%%%%%%%%%%%%%%%%%%%%%%%%%%%%%%%%%%

We first fitted one [OI] and two [SII] emission lines after subtracting the surrounding continuum 
and measured the central wavelengths 
of the [OI]$\lambda6300$, 
[SII]$\lambda6718$, [SII]$\lambda6731$ lines as shown in Table~\ref{tab:emprop}. 
The obtained redshift is $z=0.1643$. 
The measured widths (FWHMs) of the two [SII] lines are consistent with each other 
and the spectral resolution for the target is $\sim1,200$. 
This is also consistent with the fact that the seeing was about 0.5~arcsec FWHM, half of the slit size. 

Second, 
we fitted the continuum of the absorption-corrected spectrum around the H$\alpha$ region 
with a linear regression line ($\lambda$-$f_{\lambda}$) 
to isolate the AGN emission lines. 
After subtracting the continuum, we fitted broad and narrow H$\alpha$ and two [NII] emission lines 
with four Gaussians simultaneously. 
%{\bf MWR: I'm not sure if I understood the following sentence.  Please
%  check and verify that I have not made an error in my modification
%  of the sentence to clarify it.}
In this procedure, we fixed the central wavelengths 
of the two [NII] components and the narrow H$\alpha$ component,
and fixed their widths to be the same as those of the [SII] lines.
We also assume the theoretical [NII] line flux ratio 
of 2.96~(=$f$([NII]$\lambda6583$)$/f$([NII]$\lambda6548$). 
The central wavelengths of the narrow and broad H$\alpha$ are also set to be the same. 
We finally find that the width and the flux of the broad H$\alpha$ are $1880\pm120$~km~s$^{-1}$ and 
$(61.2\pm3.1)\times10^{-17}$~erg~s$^{-1}$~cm$^{-2}$, respectively. 
All these were similarly done in the previous study \citep{dong2012}. 
The emission line properties measured are summarized in Table~\ref{tab:emprop}. 

In the 300B grism spectrum (3,500-7,000\AA), 
H$\beta$ and [OIII]$\lambda4959,5007$ lines are significantly detected and 
H$\beta$ line is in general 
useful for independent BH mass and extinction measurements. 
For BH mass measurements, the width of broad H$\beta$ component 
and the monochromatic continuum luminosity at 5,100\AA\ are required. 
However, the H$\beta$ broad component is not so strong as H$\alpha$,
and the overall spectrum could be contaminated by the host galaxy continuum;
moreover, the complex FeII emission and estimation of the continuum flux
requires the detailed modeling and fitting of these two components, 
which is beyond the scope of this paper. 
Therefore, we here only measure 
the fluxes of these emission lines 
(Table~\ref{tab:emprop}) 
by summing the flux instead of fitting the lines 
to derive 
the flux ratio $f(\rm{[OIII]})/f(\rm{H}\beta)$. 
%(Table~\ref{tab:emprop}). 
%by summing the flux instead of fitting the lines. 

%\textcolor{blue}{
UV-optical variability is detected from this AGN and the clear broad H$\alpha$ 
and H$\beta$ emission lines are also detected. 
These facts indicate that the AGN UV-optical emission from the accretion disk 
does not suffer from heavy extinction. 
However, we here estimate 
possible extinction originated from dust in the host galaxy 
with the Balmer decrement method. 
As stated in the above paragraph, 
the narrow H$\beta$ emission line is not clearly detected 
and the rough lower limit of the flux ratio $f(H\alpha)/f(H\beta)>5$ 
for the narrow components is obtained. 
This flux ratio is larger than that expected in the Case B recombination and 
that of the broad components of this object ($\sim3.6$) 
which is only slightly larger than the median value of 3.06 for type-1 AGN \citep{dong2008}. 
Assuming the Calzetti extinction law \citep{calzetti2000}, we estimated 
dust extinction of 
$A(H\alpha,narrow)\gtrsim1.4$~mag (a factor of $\gtrsim3.5$) and 
$A(H\alpha,broad)\sim0.6$~mag (a factor of $\sim1.7$) 
at the H$\alpha$ wavelength, 
which corresponds to a possible BH mass underestimation by a factor of $\gtrsim1.9$ and $\sim1.3$, 
for the narrow and broad components, respectively. 
%}

%%%%%%%%%%%%%%%%%%%%%%%%%%%%%%%%%%%%%%%%%%%%%%%
\subsubsection{Black Hole Mass Measurements}
%%%%%%%%%%%%%%%%%%%%%%%%%%%%%%%%%%%%%%%%%%%%%%%

H$\alpha$ width (FWHM) and luminosity are often used for estimating BH masses,
especially for low-mass BHs 
(\cite{greene2005}; \cite{dong2012}; \cite{schramm2013}). 
We convert the broad H$\alpha$ luminosity and width 
to estimate the BH mass 
by an equation (A1) in \citet{greene2007b}, 
$M_{\rm{BH}}=(3.0^{+0.6}_{-0.5})\times10^{6}\left(L_{\rm{H\alpha}}/10^{42}~\rm{~[erg~s}^{-1}\rm{]}\right)^{0.45\pm0.03}\\
\left(\rm{FWHM_{H\alpha}}/10^{3}\rm{~[km~s}^{-1}]\right)^{2.06\pm0.06}~\rm{ [M}_\odot \rm{]}$. 
The obtained BH mass is $M_{\rm{BH}}=(2.7\pm0.6)\times10^{6}$~[M$_\odot$]. 

Monochromatic luminosity at 5100\AA, 
$\lambda L_{\lambda}(5100$\AA), and 
bolometric luminosity $\lambda L_{\rm{bol}}$ 
can be evaluated by equations in \cite{greene2005}, 
$\lambda L_{\lambda}(5100\rm{\AA})=10^{44}\left(L_{\rm{H\alpha}}/5.25\times10^{42}~\rm{[erg s}^{-1}\rm{]}\right)^{(1/1.157)}~\rm{[erg~s}^{-1}~\rm{]}$ and 
$L_{\rm{bol}}=\lambda L_{\lambda}(5100\rm{\AA})\times c_{\rm{bol}}~\rm{[erg~s}^{-1}~\rm{]}$. 
The flux density at the rest-frame 5,100\AA\  is 
found to be $4.4\times10^{-18}$~[erg~s$^{-1}$~cm$^{-2}$~\AA$^{-1}$] 
and this indicates that luminosity fraction of AGN relative to the total luminosity, 
$L_{\lambda}(\rm{5100\AA}$; AGN)$/L_{\lambda}(\rm{5100\AA}$; AGN$+$galaxy),
is 0.33 as shown in the top panel of Figure~\ref{fig:spectrum}.

\begin{figure*}
 \begin{center}
  \includegraphics[angle=-90,width=7.0cm]{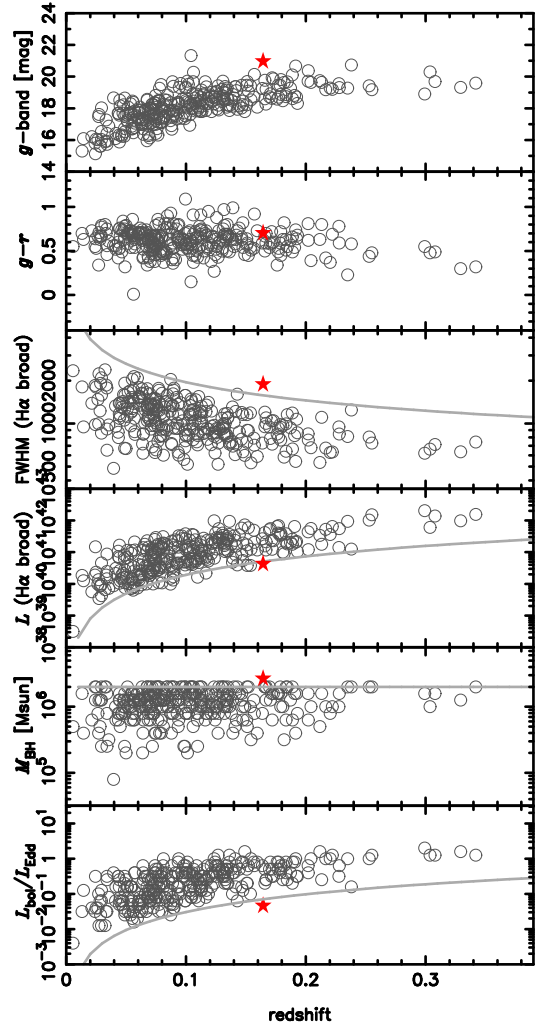} 
\end{center}
\caption{Apparent $g$-band magnitude, apparent $g-r$ color without deblending, 
FWHM and luminosity of broad H$\alpha$ emission line, 
BH mass, 
and Eddington ratio 
as a function of redshift 
of our target AGN (red star) and low-mass active BHs in \citet{dong2012} (black circle). 
The limits of the SDSS study \citep{dong2012} are shown in gray lines. 
}
\label{fig:compdong2012}
\end{figure*}

Based on the BH mass and H$\alpha$ luminosity, 
we calculate the Eddington ratio to be 0.046 by assuming a bolometric correction of $c_{\rm{bol}}=9.8$ \citep{mclure2004}. 
We locate our object in figures of 
apparent $g$-band total magnitude, 
apparent $g-r$ color without deblending, 
broad H$\alpha$ width (FWHM), 
H$\alpha$ luminosity, 
BH mass, 
and Eddington ratio, 
as a function of redshift by comparing with the sample in \cite{dong2012} 
(Figure~\ref{fig:compdong2012}). 

%{\bf MWR: I think that the flux value in the following sentence 
%     should be a lower limit -- is that right?  If I am wrong,
%     please remove my change to this sentence.}
A major limitation of the SDSS studies is 
the detection of broad H$\alpha$ emission lines, which roughly corresponds to 
a limit on the flux of broad H$\alpha$ lines, $f(\rm{H}\alpha) \geq 5.2\times10^{-16}$~[erg~s$^{-1}$~cm$^{-2}$] \citep{dong2012}. 
This can be converted to the limits of luminosity and width of H$\alpha$ ($L(\rm{H}\alpha$) and FWHM(H$\alpha$)), 
bolometric luminosity ($L_{\rm{bol}}$), and Eddington ratio ($L_{\rm{Edd}}/L_{\rm{bol}}$) 
given the definition (upper limit) of the BH masses of $2\times10^{6}$~M$_\odot$. 
These limits are shown in gray lines in Figure~\ref{fig:compdong2012}. 
Our object has a slightly higher BH mass compared to the SDSS studies 
and is close to the lower-end of the H$\alpha$ luminosity distribution, 
resulting in lower Eddington ratio, 
thanks to its apparent faintness 
relative to other galaxies in the same redshift range. 

\begin{figure*}
 \begin{center}
  \includegraphics[angle=-90,width=14cm]{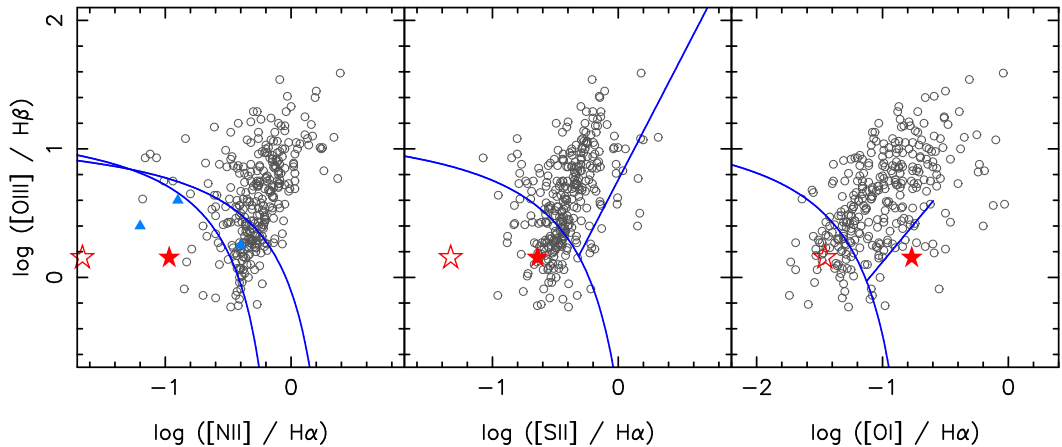} 
 \end{center}
\caption{
BPT diagram for our target (red open stars for broad and narrow components of H$\alpha$ and filled stars for only narrow H$\alpha$ emission lines) and 
objects in \cite{dong2012} (black circles) and \cite{schramm2013} (only in the left panel, blue triangles). 
Lines separating HII region, LINER, and Seyfert nuclei derived from \citet{kewley2006} are shown. 
}\label{fig:bpt}
\end{figure*}

Figure~\ref{fig:bpt} is a BPT diagram for our target and objects in \citet{dong2012}. 
We note again that 
the [OI] and [SII] emission lines can be more securely measured than the [NII] emission lines. 
We locate two points for our object in each panel. 
%\textcolor{blue}{
We also note that the flux of the H$\beta$ emission line is measured 
for the entire H$\beta$ emission line dominated by the broad component. 
%}. 
As shown in similar studies (e.g., \cite{schramm2013}), 
the location of our object indicates 
these elements are dominantly excited 
by star forming activity 
rather than the AGN activity, 
or a composite of star forming and low-ionization narrow emission-line regions (LINER)-like activities, 
%\textcolor{blue}{
if one tries to classify objects without deblending emission lines into narrow and broad components. 
%}. 
This indicates that measurements of emission line ratios do not work for selecting our object. 
On the other hand, two-thirds of the objects in the sample of \citet{dong2012} are located at the Seyfert regions as shown in Figure~\ref{fig:bpt}. 

%%%%%%%%%%%%%%%%%%%%%%%%%%%%
\subsection{Host Galaxy Properties}
\label{sec:host}
%%%%%%%%%%%%%%%%%%%%%%%%%%%%
We measure the radial profile of the host galaxy with {\it GALFIT} \citep{peng2010}. 
%We measure the radial profile of the host galaxy stellar light with GALFIT \citep{peng2010}. 
The fitting was done for the HSC images in all the epochs in $g$- and $r$-bands 
with masks covering a northern star, an eastern star, and an east-south extended source, close to the target. 
We note that the emission line contributions are small according to the spectra (Figure~\ref{fig:spectrum}). 
%Contribution from the AGN is expected to be comparable to the galaxy light contribution as shown in Figure~\ref{fig:spectrum} 
%and we 
We fit the images with a PSF at the position of the AGN (galaxy center) 
%We fit the images with a point spread function (PSF) at the position of the AGN (galaxy center) 
and an extended component with a \sersic\ profile of \sersic\ index $n$ \citep{sersic1963}. 

The fitting results are all well done (reduced $\chi^{2}\sim1$) and 
nicely consistent in all the epochs. 
A half-light radius is $r_e=3.5-4.0$~pixel ($\sim2$~kpc) and 
\sersic\ index is $n=0.5-1.0$ with measurement errors of about 0.5~pixel and 0.1, respectively. 
The $g$- and $r$-band magnitudes of the \sersic\ (extended) component are 
$21.6\pm0.2$~mag and $20.7\pm0.1$~mag, respectively, giving $g-r$ color of about 0.9. 
The axis ratios and the position angles are $0.80\pm0.05$ and $15\pm10$ degrees, respectively. 

The morphological information indicates that the host galaxy is a disk galaxy. 
Based on a recent study on galaxy morphologies by \citet{lange2015}, 
the \sersic\ index and size of the galaxy 
indicate that the host galaxy stellar mass is about $10^{9}$~M$_\odot$ 
although the intrinsic scatter is large. 
This stellar mass is within the scatter of the BH-to-total mass ratio for $2.7\times10^{6}$~M$_\odot$ objects 
shown in \citet{reines2015}. 
%In previous studies on low-mass active black holes, the host galaxies are similarly disk galaxies (?). 
In the previous SDSS studies (\cite{greene2007a}; \cite{dong2012}), 
although it is difficult to examine the morphology of the host galaxies with the image quality of the SDSS data, 
%In \cite{dong2012},  was not examined because of the SDSS image quality, 
the measured $g-r$ colors of the host galaxies are somewhat red 
but still consistent with colors of disk galaxies. 
(\cite{greene2007a}; \cite{dong2012}). 
The $g-r$ color of the host galaxy of our object is redder than the median values in those papers 
but still within the range of their $g-r$ colors. 
%\citep{kranz2010}. 

%%%%%%%%%%%%%%%%%%%%%%%%%%%%
%%%%%%%%%%%%%%%%%%%%%%%%%%%%
%%%%%%%%%%%%%%%%%%%%%%%%%%%%
\section{Discussion \& Summary}
\label{sec:summary}
%%%%%%%%%%%%%%%%%%%%%%%%%%%%
%%%%%%%%%%%%%%%%%%%%%%%%%%%%
%%%%%%%%%%%%%%%%%%%%%%%%%%%%

We selected a low-mass active BH candidate through short-time scale variability selection 
based on 1-hour-cadence imaging data with Subaru HSC, 
and spectroscopically determined that 
the object hosts an active BH,
the mass of which we find to be 
$2.7\times10^{6}$~M$_\odot$ 
based on the H$\alpha$ emission line after careful correction of the spectrum. 
This object is located at a relatively low redshift, within the redshift range 
probed by the SDSS dataset (\cite{greene2005}; \cite{dong2012}). 
Our object is relatively bright in our HSC data
and almost at the faint-end limit of the SDSS data. 
Compared with the SDSS samples, the H$\alpha$ width is broader 
and the H$\alpha$ luminosity is 
as small as those of the faintest SDSS objects 
as shown in Figure~\ref{fig:compdong2012}. 
The obtained BH mass is slightly above the upper end of those two previous studies 
(because of their mass criteria for low-mass BHs). 

Our method is observationally limited solely by variability detection. 
If we assume that variability amplitude is constant over wide ranges 
of BH mass and Eddington ratio, 
variability amplitude is expected to be proportional to AGN luminosity, which is the product of 
BH mass and Eddington ratio, as described in equation~\ref{eqn:variampl}, 
where $f_{\rm{vari}}$ and $c_{\rm{bol}}$ are variability amplitude and bolometric correction, respectively. 

\begin{eqnarray}
	\Delta{L}&=&f_{\rm{vari}}\times\frac{1}{c_{\rm{bol}}}\times
	\frac{L_{\rm{bol}}}{L_{\rm{Edd}}}\times\left(1.26\times10^{38}(M/M_\odot)\right)~\rm{[erg~s}^{-1}\rm{]}
	\label{eqn:variampl}\\
	&& {\rm where}\quad L_{\rm{Edd}}=4\pi Gcm_pM/\sigma_e=1.26\times10^{38}(M/M_\odot)~\rm{[erg~s}^{-1}\rm{]}\nonumber
\end{eqnarray}

\begin{figure*}
 \begin{center}
  \includegraphics[angle=-90,width=14cm]{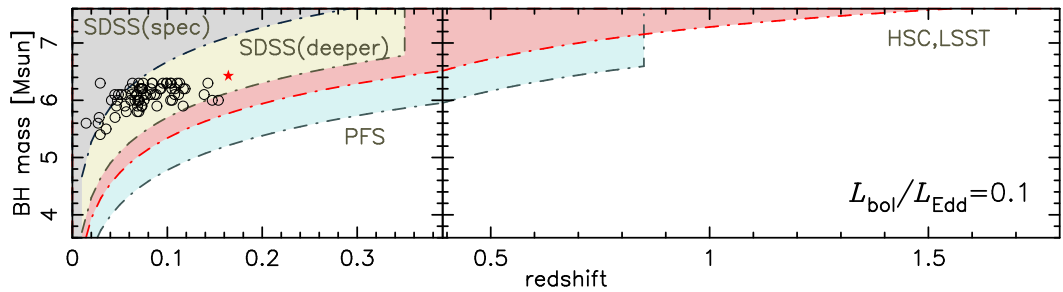} 
 \end{center}
\caption{
Detectable BH mass as a function of redshift. 
Eddington ratio ($L_{\rm{bol}}/L_{\rm{Edd}}$) of 0.1 is assumed. 
Variability amplitude of 10\% ($f_{\rm{vari}}=0.1$) is also assumed. 
%Variability amplitude ($\Delta{L}=0.1$) is also assumed. 
Our target AGN is shown as a red star and the SDSS objects with 
$L_{\rm{bol}}/L_{\rm{Edd}}\sim0.1$ are plotted in open circles. 
}
\label{fig:expectation}
\end{figure*}

By roughly assuming $f_{\rm{vari}}=0.1$~(10\%) \citep{vandenberk2004}, 
variability amplitude defined as magnitude of differential flux in subtracted images\footnote{This is a value directly measured from observations,
whereas estimation of total AGN flux requires AGN+galaxy decomposition with a certain method.}
can be calculated as a function of redshift. 
For a low-mass active BH at $z=0.164$, the total flux including the host galaxy component is about 15~$\mu$Jy 
and the AGN component is a third of the total flux: about 5~$\mu$Jy. 
The variability amplitude from hours to a day is about 10\%,
while that at time scales longer than a day is larger than 20\%. 
We here assume that the power-law index of AGN spectra ($f_{\nu}\propto\nu^{\alpha_{\nu}}$) is $\alpha_{\nu}=-0.44$ \citep{vandenberk2001}. 
We also assume $g$-band HSC observations with 
a typical depth in a 10~minute exposure of $g\sim26$~mag. 
In this case, 
variability detection can go down to $g_{\rm{vari}}\sim25.6$~mag, 
which is shallower than the original images by a factor of $\sqrt{2}$. 
This is because we need to investigate intranight variability, not variability relative to deep reference images. 
Under these assumptions, we can calculate the BH mass range from which we can detect variability as shown in Figure~\ref{fig:expectation}. 

In the SDSS studies, the optical wavelength range 
and depths ($r<17.77$ for the spectroscopic sample 
described in \cite{strauss2002} 
and $r\sim20$ for some deeper specific samples) 
limit the redshift ($z<0.352$ for H$\alpha$; \cite{greene2007b}; \cite{dong2012}) and brightness ranges. 
Figure~\ref{fig:expectation} indicates that 
our variability survey can find smaller active BHs at higher redshifts than the SDSS 
although AGN variability amplitude is 
not so large that it is more difficult to detect the variability than the objects themselves. 
The Prime Focus Spectrograph (PFS; \cite{takada2014}), to be attached to the 8.2-m Subaru telescope in near future, will achieve 
a wider wavelength coverage, up to $J$-band, 1.26~$\mu$m, corresponding to the limit of H$\alpha$ detection at $z\sim0.85$, 
and deeper spectroscopic capability. 
Compared to these two large spectroscopic surveys, which naively require blind galaxy spectroscopic observations, 
our selection based on multi-epoch imaging data over one or a few nights is less expensive. 
Since in the blind spectroscopic SDSS surveys, 
the low-mass active BH ($<2\times10^6$~M$_\odot$) fraction is as small as 0.03\%~($174/544,127$; \cite{greene2007a}; \cite{greene2007b}) 
and 0.07\%~(309/451,000; \cite{dong2012}), 
our HSC survey has the advantage of effectively selecting low-mass active BH candidates as spectroscopic targets. 
%\textcolor{blue}{
For example, given these low success rates of the blind spectroscopic surveys 
and the number of fibers of PFS (2,400), 
one fiber is accidentally assigned to a low-mass active BH on average but 
the expected number density of these populations is larger 
based on the number density shown in the previous SDSS study \citep{greene2007b}. 
%}
%\textcolor{blue}{
In our current dataset, we have photometrically detected short timescale variability from 
an order of ten candidates per HSC field-of-view, whose nature should be examined with spectroscopic data.
%}

The sample size will be enhanced by larger HSC surveys and coming LSST survey.  
%%{\bf MWR: I do not understand what the following sentence means.  I have not modified it at all.}
%%On the other hand, variability amplitude is 
%%not so large that detection of variability is more difficult to detect objects themselves. 
%On the other hand, variability amplitude is 
%not large, so it is more difficult to detect the variability than the objects themselves. 
%Figure~\ref{fig:expectation} indicates that 
%our variability survey can find smaller active BHs at higher redshifts than the SDSS. 
Low-mass active BH candidates selected based on our method will be good candidates for 
systematic spectroscopic observations with PFS. 
This variability method basically has no redshift limitation except for the effect of Lyman break and brightness,
although AGN identification and BH mass measurements in spectroscopic observations with current facilities 
might limit the capability of our method. This will be improved in the era of 30-m class telescopes. 

%%%%%%%%%%%%%%%%%%%%%%%%%%%%%%%%%%%%%%%

\begin{ack}
%This research was in part supported by Grants-in-Aid for Scientific Research of JSPS (, 23740157, 24740117, , 26400222, , ), MEXT (25103515, 15H00788), the World Premier %International Research Center Initiative, MEXT, Japan, the research grant program of Toyota foundation (D11-R-0830), and the RFBR-JSPS bilateral program.
This research was partially supported by 
JSPS Grants-in-Aid for Scientific Research (15H05440, 25800103, 15H02075, 23224004, and 26400222), 
the Toyota Foundation (D11-R-0830), and 
the World Premier International Research Center Initiative, MEXT, Japan.
%The work of SB is supported by the Russian Science Foundation grant 14-12-00203. 
The work of SB is supported by the RF grant 11.G34.31.0047.
This paper is based in part on data collected at Subaru Telescope, 
which is operated by the National Astronomical Observatory of Japan.
This paper makes use of software developed for the Large Synoptic Survey Telescope. 
We thank the LSST Project for making their code available as free software at http://dm.lsstcorp.org. 
We also appreciate a kind help by Prof. Michael W. Richmond for improving the
English grammar of the manuscript.
\end{ack}

%%%
% See the manual for the detail.
%%%

%\clearpage

%\rotate
\begin{table*}
  \tbl{Light Curves Relative to the Reference Images Taken on May 24, 2015.}{%
  \begin{tabular}{ccrccrrrrr}%l}
      \hline
date & MJD & Day & Inst & filter & $t_{\rm{exp}}$ & flux(sub) & mag(sub) & seeing\\
 &  &  &  &  & [sec] & [$\mu$Jy] & [mag] & [arcsec]\\\hline
2014-07-02 & 56840.299 & 1   & HSC   & $g$ &  600 & $-3.756\pm0.036$ & $22.46^{+0.01}_{-0.01}$ & 0.54\\
2014-07-02 & 56840.342 & 1   & HSC   & $g$ &  600 & $-3.787\pm0.029$ & $22.45^{+0.01}_{-0.01}$ & 0.75\\
2014-07-02 & 56840.526 & 1   & HSC   & $g$ &  600 & $-3.343\pm0.036$ & $22.59^{+0.01}_{-0.01}$ & 1.19\\
2014-07-03 & 56841.293 & 2   & HSC   & $g$ &  600 & $-2.583\pm0.040$ & $22.87^{+0.02}_{-0.02}$ & 0.71\\
2014-07-03 & 56841.338 & 2   & HSC   & $g$ &  600 & $-2.668\pm0.033$ & $22.83^{+0.01}_{-0.01}$ & 0.70\\
2014-07-03 & 56841.500 & 2   & HSC   & $g$ &  600 & $-2.756\pm0.037$ & $22.80^{+0.01}_{-0.01}$ & 0.79\\
2015-05-24 & 57166.467 & 327 & HSC   & $g$ & 1080 & $0$              & -                       & 0.62\\
2015-06-22 & 57195.387 & 356 & FOCAS & $g$ &  300 & $-2.881\pm0.103$ & $22.75^{+0.04}_{-0.04}$ & 0.61\\
\hline
2014-07-02 & 56840.442 & 1   & HSC & $r$ &  600 & $-3.782\pm0.064$ & $22.46^{+0.02}_{-0.02}$ & 0.72\\
2014-07-03 & 56841.422 & 2   & HSC & $r$ &  600 & $-2.708\pm0.059$ & $22.82^{+0.02}_{-0.02}$ & 0.49\\
2015-05-24 & 57166.398 & 327 & HSC & $r$ &  360 & $0$              & -                       & 0.68\\
%
%% 
%%%%%%%%%%%%%%%%%
      \hline
    \end{tabular}}\label{tab:lc}
\begin{tabnote}
%This is table note.
\end{tabnote}
\end{table*}

\begin{table*}
  \tbl{Emission Line Properties.}{%
  \begin{tabular}{lrrr}
      \hline
      Line & $\lambda_{\rm{cen}}$ [\AA] & Width [\AA] & Flux [$10^{-17}$~erg s$^{-1}$ cm$^{-2}$] \\ 
      \hline
$[$SII$]$$\lambda6731$  & $7819.99$ & $7.69\pm1.72$ & $1.70\pm0.33$\\  % [SII]a
$[$SII$]$$\lambda6718$  & $7836.76$ & $5.55\pm0.78$ & $1.88\pm0.27$\\  % [SII]b
$[$NII$]$$\lambda6584$  & -- & -- & $0.19\pm0.82$\\ % [NII]b
H$\alpha$ (narrow)      & -- & -- & $15.69\pm0.95$\\ % Halpha, narrow
H$\alpha$ (broad)       & -- & $47.95\pm2.97$ & $61.16\pm3.13$\\ % Halpha, broad
$[$NII$]$$\lambda6548$  & -- & -- & $0.57\pm0.28$\\ % [NII]a
$[$OI$]$$\lambda6300$   & $7338.93$ & $11.72\pm1.92$ & $2.68\pm0.40$\\  % [OI]
$[$OIII$]$$\lambda5007$ & -- & -- & $11.92\pm2.26$\\ % [OIII]b
$[$OIII$]$$\lambda4959$ & -- & -- & $4.00\pm0.76$\\ % [OIII]a
H$\beta$                & -- & -- & $17.16\pm0.93$\\ % Hbeta, broad
      \hline
    \end{tabular}}\label{tab:emprop}
\begin{tabnote}
We show only fitted parameters. 
The redshift determined by the [OI] and [SII] emission line wavelengths. 
\end{tabnote}
\end{table*}

\end{document}